\documentclass[preprint,11pt,3p]{elsarticle}

\usepackage{amssymb}

\usepackage{amsthm}
\usepackage{amsmath}

\newcommand{\curl}[1]{\nabla \times #1} 
\newcommand*\diff{\mathop{}\!\mathrm{d}}

\renewcommand{\vec}[1]{\mathbf{#1}}

\usepackage[figuresright]{rotating}

\usepackage{caption}
\usepackage{subcaption}

\journal{JQSRT}

\begin{document}

\begin{frontmatter}


\title{Photonic Nanojets in Optical Tweezers}


\author{Antonio Alvaro Ranha Neves}
\ead{antonio.neves@ufabc.edu.br}
\address{Centro de Ci\^encias Naturais e Humanas, Universidade Federal do ABC (UFABC), Santo Andr\'e - S\~ao Paulo, 09.210-170, Brazil}

\begin{abstract}
Photonic nanojets have been brought into attention ten years ago for potential application in ultramicroscopy, because of its sub-wavelength resolution that can enhance detection and interaction with matter. For these novel applications under development, the optical trapping of a sphere acts as an ideal framework to employ photonic nanojets. In the present study, we generated nanojets by using a highly focused incident beam, in contrast to traditional plane waves. The method inherits the advantage of optical trapping, especially for intracellular applications, with the microsphere in equilibrium on the beam propagation axis and positioned arbitrarily in space. Moreover, owing to optical scattering forces, when the sphere is in equilibrium, its center shifts with respect to the focal point of the incident beam. However, when the system is in stable equilibrium with a configuration involving optical tweezers, photonic nanojets cannot be formed. To overcome this issue, we employed double optical tweezers in an unorthodox configuration involving two collinear and co-propagating beams, the precise positioning of which would turn on/off the photonic nanojets, thereby improving the applicability of photonic nanojets.
\end{abstract}


\end{frontmatter}


\section{Introduction}
\label{sec:Introduction}

A photonic nanojet (PNJ) is a narrow (subwavelength) and elongated region with high intensity located at the shadow-side surface of an illuminated loss-less dielectric microcylinder or microsphere \cite{Heifetz2009,Kim2011}. In the present study, we focus on PNJs generated by a dielectric microsphere acting as a focusing lens, which results in the build up of spherically aberrated rays in the focal region. This build up of rays in the spatially localized high-energy-density region constitute the external caustic \cite{Chen2004,Lock2000}. Therefore, PNJs are not a new phenomenon, but they have attracted renewed interest because of recent technological advances allowing the exploitation of this high-energy-density region. Nonetheless, contrary to a possible understanding \textit{via} geometrical optics or catastrophe theory, we aim to obtain an exact solution of Maxwell's equations by employing the generalized Lorenz-Mie theory (GLMT) \cite{Gouesbet2011}, as in the study by Devilez \textit{et al.} \cite{Devilez2009}. Such focusing of light using a microsphere has also been investigated by Kofler \textit{et al.} \cite{Kofler2006} using the uniform caustic asymptotic method, to obtain analytical expressions for the intensity by matching the geometrical-optic solutions with Bessoid integrals. However, for optical trapping applications, the scattering object is a microsphere of size comparable to the illuminating wavelength and consequently outside the regime of geometrical optics.

Optical trapping results from the change in linear momentum of a beam scattered by a microsphere, producing a resultant restoring force. The most common setup for laser trapping is of optical tweezers \cite{Dholakia2006,Jonas2008}. PNJs and optical tweezers (OT) have attracted attention because, in combination, they can be highly useful in applications including nanoscale processing \cite{Munzer2001,Piglmayer2002,McLeod2008,Pereira2008,Zhang2013,David2014}, high-resolution microscopy \cite{Chen2004,Li2005,Lee2013,Darafsheh2014,Ye2014}, and enhanced inelastic spectroscopy, such as Raman scattering \cite{Yi2007,Kasim2008,Cardenas2013,Alessandri2014}, coherent anti-Stokes Raman
scattering \cite{Upputuri2014,Huang2014} and fluorescence \cite{Gerard2008,Yang2014}, or elastic enhancement through backscattering from nanoparticles \cite{Chen2006}. A recent study that aimed to position PNJs in a controlled manner involved a microsphere attached to a movable micropipette \cite{Krivitsky2013} and magnetically controlled Janus spheres \cite{Li2014}. In addition, an optically trapped non-spherical dielectric particle in a Bessel beam was employed to generate PNJs from another pulsed laser for direct laser writing \cite{Tsai2013}. The disadvantage of this approach is in the engineering of particular non-spherical particles, each exhibiting a characteristic PNJ. Other studies on optical forces and PNJs are related to the radiation pressure effect produced by a PNJ on nearby nanoparticles \cite{Cui2008,Valero2012,Valero2013}. In contrast, the scope of this study is different; we investigate the PNJs generated from an optically trapped microsphere in terms of the incoming beam, trading the complexity of structuring special shaped particles for the ease of manipulating the trapping beam.

It is widely known that the main features of PNJs are waists smaller than the diffraction limit and propagation over several wavelengths without significant diffraction. These features lead to the potential application of PNJs for developing spectroscopic methods with high spatial resolution and high detection sensitivity through backscattering enhancement. Therefore, the characteristics of a nanojet have been described with a few parameters: the location of the intensity maximum, the distance from this maximum to the sphere surface (radial shift), the distance from the intensity peak to the point where the intensity decays to $e^{-2}$ times the initial intensity in the direction of the beam (decay length), and the PNJ width, which is $e^{-2}$ waist of the jet \cite{Itagi2005}. The behavior of PNJs for a given wavelength is as follows: the PNJ waist widening is proportional to the relative refractive index, and the PNJ lengthening is due to the increasing sphere size. This transverse confinement (PNJ waist) has been approximated as a Gaussian, while a part of the longitudinal confinement (decay length) is approximated as a Lorentzian profile \cite{Devilez2009spie}. In the present work, it is shown that a bessoidal-type surface best describes the shape of light confined by these PNJs originating from the interference between the scattered field and the incident beam \cite{Devilez2009} and therefore the extinction component of the Poynting vector. 

The modeling of PNJs originating from plane-wave incidence involves only four parameters: the refractive indices of the particle and its surroundings, wavelength of the incident wave, and radius of the sphere. However, with an incident focused beam, which is required for an optical trap, the complexity of the modeling increases with additional parameters: the numerical aperture and location of the waist with respect to the scatterer. The positioning of the incident focused beam with respect to the scatterer is very important, and we demonstrate that it is responsible for switching the PNJs on and off for an optically trapped microsphere.

\section{Theoretical model of photonic nanojets}
\label{sec:TheoreticalModel}

Any Maxwellian beam can be expressed by its electromagnetic field in terms of partial waves. We start with the general expression for the electromagnetic fields of interest: the incident and scattered fields. Later, we simplify this to a two-dimensional problem, which is useful for a 2D plot and subsequently for a 1D profile such as the longitudinal intensity of the PNJ.

\subsection{Generalized Lorenz-Mie theory}
\label{sec:GLMT}
The basic idea of GLMT is that a beam, which is expressed by a solution to Maxwell's equations (i.e. Maxwellian beam), can be written as an infinite series of spherical functions and spherical harmonics, each multiplied by a coefficient called a beam shape coefficient (BSC). These BSCs completely describe the incident, internal, and scattered electromagnetic fields in terms of partial-waves series. The correct determination of these radially independent BSCs allows for the precise determination of the observed electromagnetic phenomena \cite{Gouesbet2011}.

The present notation follows that of previous works \cite{Neves2006ol,Neves2006oe,Neves2007}, expressing the electromagnetic fields in terms of spherical vector wave functions even though the original formulation of GLMT was in the framework of the Bromwich scalar functions \cite{Gouesbet1999,Gouesbet2000}. The microsphere scatterer has a radius $a$ and real refractive index $n$, and the incident beam is directed towards the positive $z$ axis of the rectangular coordinate system. The incident wave has a time dependence $exp(-i \omega t)$ (omitted for clarity), wavelength $\lambda$, and wave number $k=2\pi/\lambda$. The scatterer and the surrounding medium are homogeneous, isotropic, and non-magnetic.  The incident and scattered electromagnetic fields can be written as follows:

\begin{equation}
\vec{E}_{inc}=E_0\sum_{p=1}^{\infty}\sum_{q=-p}^{p} G_{pq}^{TM} \vec{N}_{pq}(\vec{r}) + G_{pq}^{TE} \vec{M}_{pq}(\vec{r}),
\label{eq:Einc}
\end{equation}

\begin{equation}
\vec{H}_{inc}=H_0\sum_{p=1}^{\infty}\sum_{q=-p}^{p}G_{pq}^{TM}\vec{M}_{pq}(\vec{r})- G_{pq}^{TE} \vec{N}_{pq}(\vec{r}),
\label{eq:Hinc}
\end{equation}

\begin{equation}
\vec{E}_{sca}=E_0\sum_{p=1}^{\infty}\sum_{q=-p}^{p} a_{pq} \vec{N}_{pq}(\vec{r}) + b_{pq} \vec{M}_{pq}(\vec{r}),
\label{eq:Esca}
\end{equation}

\begin{equation}
\vec{H}_{sca}=H_0\sum_{p=1}^{\infty}\sum_{q=-p}^{p} a_{pq} \vec{M}_{pq}(\vec{r})- b_{pq} \vec{N}_{pq}(\vec{r}),
\label{eq:Hsca}
\end{equation}

where the terms involving the spherical functions and vectorial spherical harmonics can be abbreviated as

\begin{equation}
\vec{N}_{pq}(\vec{r})=\frac{i}{k} \curl{\vec{M}_{pq}(\vec{r})},
\label{eq:N}
\end{equation}

\begin{equation}
\vec{M}_{pq}(\vec{r})=z_{pq}(kr) \vec{X}_{pq}(\theta,\phi),
\label{eq:M}
\end{equation}

where $z_{pq}(kr)$ denotes spherical Bessel or spherical Hankel functions, depending on whether it expresses the incident or scattered field, respectively. The vectorial spherical harmonics are defined as $\vec{X}_{pq}(\vec{r})=\vec{L} Y_{lm}(\vec{r})/\sqrt{l(l+1)}$, where $Y_{lm}(\vec{r})$ denotes the scalar spherical harmonics and $\vec{L}=-i \vec{r}\times d/d\vec{r}$ is the angular momentum operator in direct space. The scattering partial-wave coefficients $a_{pq}$ and $b_{pq}$ are related to the Mie scattering coefficients by

\begin{equation}
-a_p=\frac{a_{pq}}{G_{pq}^{TM}}=\frac{m \psi_p(mx)\psi_p'(x)-\psi_p'(mx)\psi_p(x)}{\psi_p'(mx)\xi_p(x)-m \psi_p(mx)\xi_p'(x)},
\end{equation}

\begin{equation}
-b_p=\frac{b_{pq}}{G_{pq}^{TE}}=\frac{m \psi_p'(mx)\psi_p(x)-\psi_p(mx)\psi_p'(x)}{\psi_p(mx)\xi_p'(x)-m \psi_p'(mx)\xi_p(x)}.
\end{equation}

The minus sign was included to adopt the more frequent notation for the Mie coefficients $a_{p}$ and $b_{p}$. The coefficients $G_{pq}^{TM}$ and $G_{pq}^{TE}$ represent the expansion coefficients (BSCs) of the incident fields. The BSCs are defined as follows:

\begin{equation}
G_{pq}^{TM}=-\frac{g_p}{E_0} \int_{0}^{\pi} d\theta \sin\theta \int_{0}^{2\pi} d\phi Y_{pq}^*(\theta,\phi) \vec{E} \cdot \hat{r},
\end{equation}

\begin{equation}
G_{pq}^{TE}=\frac{g_p}{H_0} \int_{0}^{\pi} d\theta \sin\theta \int_{0}^{2\pi} d\phi Y_{pq}^*(\theta,\phi) \vec{H} \cdot \hat{r},
\end{equation}

where $g_p=kr/(j_p(kr) \sqrt{n(n+1)})$ was introduced to shorten the equation. Note that the explicit cancellation of the radial dependence in $g_p$ has been the basis of various approximation techniques for the BSC \cite{Gouesbet2011,Neves2006ol}. Recently, the analytical solution for any type of Maxwellian beam has been demonstrated \cite{Moreira2010}.

For all the incident beams described henceforth, the axis of symmetry coincides with that of the sphere, and the scenario is termed on-axis. This is due to the fact that the equilibrium position for a homogeneous sphere lies on the axis of symmetry. For this particular configuration, the double sum in the partial wave expansion, Eqs. \eqref{eq:Einc}-\eqref{eq:Hsca}, simplifies because of contributions exclusively from $q=\pm 1$ \cite{Gouesbet1996,Neves2007}. For this case, it would be best to introduce a new vectorial function to describe our electromagnetic fields:

\begin{equation}
\vec{N}_{p}^{\pm}(\vec{r})=\vec{N}_{p,1}(\vec{r})\pm\vec{N}_{p,-1}(\vec{r}),
\label{eq:Naxis}
\end{equation}

\begin{equation}
\vec{M}_{p}^{\pm}(\vec{r})=\vec{M}_{p,1}(\vec{r})\pm\vec{M}_{p,-1}(\vec{r}),
\label{eq:Maxis}
\end{equation}

thereby simplifying the analytical solution up to this point using the symmetrical relations for $q=\pm 1$. Further simplification of the BSC would require knowledge of the exact vectorial electromagnetic fields near the focus, as will be shown in section \ref{sec:Approx2D}.

\subsection{Plane wave incidence}
\label{sec:PlaneW}
Any incident field can be expressed as an infinite series of vectorial spherical harmonics. As an initial approach, an arbitrary polarized plane wave travelling along the z-axis ($\vec{k}=k\hat{z}$) is used to validate the results and compare them with known results. The electric field is

\begin{equation}
\vec{E}=E_0 e^{i\vec{k}\cdot\vec{r}} \left( p_{x}, p_{y}, 0 \right),
\end{equation}

where the polarization is described by the components of the Jones vector, $(p_x,p_y)$. The magnetic field can be easily determined from the relation

\begin{equation}
\vec{H}=\frac{\vec{k}}{\omega \mu} \times \vec{E}=H_0 e^{i\vec{k}\cdot\vec{r}} \left( -p_{y}, p_{x}, 0 \right).
\end{equation}

The radial components of the fields in spherical coordinates are

\begin{equation}
\vec{E} \cdot \hat{r} = E_0 e^{i k r \cos\theta} \sin\theta \left( p_{x} \cos\phi + p_{y} \sin\phi \right),
\end{equation}
\begin{equation}
\vec{H} \cdot \hat{r} = H_0 e^{i k r \cos\theta} \sin\theta \left( p_{x} \sin\phi - p_{y} \cos\phi \right),
\end{equation}

where we used $\hat{r}=\sin\theta\cos\phi\hat{x}+\sin\theta\sin\phi\hat{y}+\cos\theta\hat{z}$. The integrals over the solid angle are presented in \ref{Sec:AppIntegrals}, and it can be seen that only the $q=\pm1$ components are present. The plane-wave (pw) BSCs are

\begin{equation}
G_{p,\pm 1}^{TM,pw}=G_p^{pw} \left( \mp i p_{x} - p_{y} \right), \quad G_{p,\pm 1}^{TE,pw}=G_p^{pw} \left( p_{x} \mp i p_{y} \right),
\end{equation}

\begin{equation}
G_{p}^{pw}=\sqrt{\pi (2p+1)} i^{p}.
\end{equation}

\subsection{Description of highly focused beam}
\label{sec:BeamDes}

The commonly employed Davis description of Gaussian beams, a perturbation method that introduces higher-order corrections to the paraxial approximation, yields a condition that is rarely met for strong focusing (as in optical tweezers). Consequently there is a need to go beyond a paraxial-expansion description. To achieve this aim, we resort to an early description based on the angular spectrum representation, which is later applied to optical trapping to result in a solution for a completely arbitrary vectorial Maxwellian beam \cite{Neves2006ol}. This new approach has the benefit of providing an analytical expression for the BSCs as a function of beam position and polarization directly with respect to the sphere without the need to rely on vector translation theorems requiring high computational cost. Results obtained by using this approach have recently been confirmed independently through an aberrated photothermal microscopy measurement to be in excellent agreement with the theory \cite{Selmke2012}.

To present a complete study of PNJs in optical tweezers, we start with a description of the highly focused incident beam in the framework of GLMT. The on-axis case is considered here, since the equilibrium trap position for a homogeneous sphere lies on the propagation axis. The exact BSCs for an on-axis highly focused beam (fb) positioned at $z_0$ with respect to the sphere center are \cite{Neves2007}

\begin{equation}
G_{p,\pm 1}^{TM,fb}=G_p^{fb} \left( i p_{x} \pm  p_{y} \right), \quad G_{p,\pm 1}^{TE,fb}=G_p^{fb} \left(\mp p_{x} +i p_{y} \right),
\end{equation}

\begin{equation}
G_p^{fb}=i kf e^{ikf} \frac{G_p^{pw}}{p(p+1)} \int_{0}^{\alpha_{max}} \diff\alpha \sin\alpha \sqrt{\cos\alpha} e^{ikz_0\cos\alpha} e^{-(f \sin\alpha/\omega)^2}\left[\pi_p^1(\alpha)+ \tau_p^1(\alpha) \right].
\label{eq:BSCfb}
\end{equation}

In terms of the two angular functions, which involves associated Legendre functions,

\begin{equation}
\pi_p^q(\theta)=\frac{P_p^q(\cos\theta)}{\sin\theta}, \qquad \tau_p^q(\theta)=\frac{d}{d \theta} P_p^q(\cos\theta).
\end{equation}

\subsection{Approximations for fields along the incident plane (2D)}
\label{sec:Approx2D}
When plotting the fields in 2D, in the $x-z$ plane, we will take $\phi=0$ but $\theta \ne 0$ in Eqs. \eqref{eq:Naxis}-\eqref{eq:Maxis}. The vectorial incident and scattered fields from Eqs. \eqref{eq:Einc}-\eqref{eq:Hsca} becomes for the on-axis case,

\begin{align}
\frac{\vec{E_{inc}}}{E_0} &= -2 i \sum_{p=1}^{\infty} \frac{G_p}{p(p+1)} \sqrt{\frac{(2p+1)}{4 \pi}} \biggl\{ \frac{z_p(kr)}{kr}
\bigg[ \frac{1}{\sin\theta} \left( \tau_p^1(\theta) \cos\theta - \pi_p^1(\theta) \right) - \frac{\partial \tau_p^1(\theta)}{\partial \theta} \bigg] p_{x} \hat{r} \nonumber \\
&-\left[ \left((p+1)\frac{z_p(kr)}{kr}-z_{p+1}\right) \tau_p^1(\theta) -iz_p(kr) \pi_p^1(\theta) \right] p_{x} \hat{\theta} -\left[ \left((p+1)\frac{z_p(kr)}{kr}-z_{p+1}\right) \pi_p^1(\theta) -iz_p(kr) \tau_p^1(\theta) \right] p_{y} \hat{\phi}
\biggl\},
\end{align}

\begin{align}
\frac{\vec{H_{inc}}}{H_0}&=2 i \sum_{p=1}^{\infty} \frac{G_p}{p(p+1)} \sqrt{\frac{(2p+1)}{4 \pi}} \biggl\{ \frac{z_p(kr)}{kr} 
\bigg[ \frac{1}{\sin\theta} \left( \tau_p^1(\theta) \cos\theta - \pi_p^1(\theta) \right) - \frac{\partial \tau_p^1(\theta)}{\partial \theta} \bigg] p_{y} \hat{r} \nonumber \\
&-\biggl[\left((p+1)\frac{z_p(kr)}{kr} - z_{p+1}(kr)\right)\tau_p^1(\theta) +i z_p(kr) \pi_p^1(\theta) \biggl] p_{y} \hat{\theta} +\biggl[\left((p+1)\frac{z_p(kr)}{kr} - z_{p+1}(kr)\right)\pi_p^1(\theta) +i z_p(kr) \tau_p^1(\theta)\biggl] p_{x} \hat{\phi} \biggl\},
\end{align}

\begin{align}
\frac{\vec{E_{sca}}}{E_0}&=2i \sum_{p=1}^{\infty} \frac{G_p}{p(p+1)} \sqrt{\frac{(2p+1)}{4 \pi}} \biggl\{ a_p \frac{z_p(kr)}{kr} \bigg[ \frac{1}{\sin\theta} \left( \tau_p^1(\theta) \cos\theta - \pi_p^1(\theta) \right) - \frac{\partial \tau_p^1(\theta)}{\partial \theta} \bigg] p_x \hat{r} \nonumber \\
&-\bigg[a_p \left((p+1)\frac{z_p(kr)}{kr} - z_{p+1}(kr)\right) \tau_p^1(\theta) +i b_p z_p(kr) \pi_p^1 \bigg] p_x \hat{\theta} -\bigg[a_p \left((p+1)\frac{z_p(kr)}{kr} - z_{p+1}(kr)\right) \pi^1(\theta) +i b_p z_p(kr) \tau_p^1 \bigg] p_y \hat{\phi} \biggl\},
\end{align}

\begin{align}
\frac{\vec{H_{sca}}}{H_0}&=2i \sum_{p=1}^{\infty} \frac{G_p}{p(p+1)} \sqrt{\frac{(2p+1)}{4 \pi}} \biggl\{ -b_p \frac{z_p(kr)}{kr} \bigg[ \frac{1}{\sin\theta} \left( \tau_p^1(\theta) \cos\theta - \pi_p^1(\theta) \right) - \frac{\partial \tau_p^1(\theta)}{\partial \theta} \bigg] p_y \hat{r} \nonumber \\
&+\bigg[b_p \left((p+1)\frac{z_p(kr)}{kr} - z_{p+1}(kr)\right) \tau_p^1(\theta) +i a_p z_p(kr) \pi_p^1 \bigg] p_y \hat{\theta} -\bigg[b_p \left((p+1)\frac{z_p(kr)}{kr} - z_{p+1}(kr)\right) \pi^1(\theta) +i a_p z_p(kr) \tau_p^1 \bigg] p_x \hat{\phi} \biggl\}.
\end{align}
 
For the plane-wave BSCs, (i.e. $G_p=G_p^{pw}$), results in

\begin{align}
\frac{\vec{E_{inc}}}{E_0} &= -i \sum_{p=1}^{\infty} \frac{(2p+1)}{p(p+1)} i^p \biggl\{ \frac{j_p(kr)}{kr}
\bigg[ \frac{1}{\sin\theta} \left( \tau_p^1(\theta) \cos\theta - \pi_p^1(\theta) \right) - \frac{\partial \tau_p^1(\theta)}{\partial \theta} \bigg] p_{x} \hat{r} \nonumber \\
&-\left[ \left((p+1)\frac{j_p(kr)}{kr}-j_{p+1}\right) \tau_p^1(\theta) -ij_p(kr) \pi_p^1(\theta) \right] p_{x} \hat{\theta} -\left[ \left((p+1)\frac{j_p(kr)}{kr}-j_{p+1}\right) \pi_p^1(\theta) -ij_p(kr) \tau_p^1(\theta) \right] p_{y} \hat{\phi}
\biggl\},
\end{align}

\begin{align}
\frac{\vec{H_{inc}}}{H_0}&=i \sum_{p=1}^{\infty} \frac{(2p+1)}{p(p+1)} i^p \biggl\{ \frac{j_p(kr)}{kr} 
\bigg[ \frac{1}{\sin\theta} \left( \tau_p^1(\theta) \cos\theta - \pi_p^1(\theta) \right) - \frac{\partial \tau_p^1(\theta)}{\partial \theta} \bigg] p_{y} \hat{r} \nonumber \\
&-\biggl[\left((p+1)\frac{j_p(kr)}{kr} - j_{p+1}(kr)\right)\tau_p^1(\theta) +i j_p(kr) \pi_p^1(\theta) \biggl] p_{y} \hat{\theta} +\biggl[\left((p+1)\frac{j_p(kr)}{kr} - j_{p+1}(kr)\right)\pi_p^1(\theta) +i j_p(kr) \tau_p^1(\theta)\biggl] p_{x} \hat{\phi} \biggl\},
\end{align}

\begin{align}
\frac{\vec{E_{sca}}}{E_0}&=i \sum_{p=1}^{\infty} \frac{(2p+1)}{p(p+1)} i^p \biggl\{ a_p \frac{h_p^{(1)}(kr)}{kr} \bigg[ \frac{1}{\sin\theta} \left( \tau_p^1(\theta) \cos\theta - \pi_p^1(\theta) \right) - \frac{\partial \tau_p^1(\theta)}{\partial \theta} \bigg] p_x \hat{r} \nonumber \\
&-\bigg[a_p \left((p+1)\frac{h_p^{(1)}(kr)}{kr} - h_{p+1}^{(1)}(kr)\right) \tau_p^1(\theta) +i b_p h_p^{(1)}(kr) \pi_p^1 \bigg] p_x \hat{\theta} -\bigg[a_p \left((p+1)\frac{h_p^{(1)}(kr)}{kr} - h_{p+1}^{(1)}(kr)\right) \pi^1(\theta) +i b_p h_p^{(1)}(kr) \tau_p^1 \bigg] p_y \hat{\phi} \biggl\},
\end{align}

\begin{align}
\frac{\vec{H_{sca}}}{H_0}&=i \sum_{p=1}^{\infty} \frac{(2p+1)}{p(p+1)} i^p \biggl\{ -b_p \frac{h_p^{(1)}(kr)}{kr} \bigg[ \frac{1}{\sin\theta} \left( \tau_p^1(\theta) \cos\theta - \pi_p^1(\theta) \right) - \frac{\partial \tau_p^1(\theta)}{\partial \theta} \bigg] p_y \hat{r} \nonumber \\
&+\bigg[b_p \left((p+1)\frac{h_p^{(1)}(kr)}{kr} - h_{p+1}^{(1)}(kr)\right) \tau_p^1(\theta) +i a_p h_p^{(1)}(kr) \pi_p^1 \bigg] p_y \hat{\theta} -\bigg[b_p \left((p+1)\frac{h_p^{(1)}(kr)}{kr} - h_{p+1}^{(1)}(kr)\right) \pi^1(\theta) +i a_p h_p^{(1)}(kr) \tau_p^1 \bigg] p_x \hat{\phi} \biggl\}.
\end{align}

Note that for the Poynting vector in the $x-z$ plane, we only need the $r$ and $\theta$ components. We are thus interested in the following products from the vectorial fields above:

\begin{equation}
S_r \hat{r} + S_\theta \hat{\theta} = \hat{r} \left(E_\theta H_\phi - E_\phi H_\theta\right) + \hat{\theta} \left(E_\phi H_r - E_r H_\phi\right).
\end{equation}

For the particular 2D case, since the photonic nanojet is extremely close to the z-axis immediately outside the spherical scatterer, the small-angle approximation for small $\theta$ is justified:

\begin{equation}
P_p^q(\cos\theta) \approx (-1)^q \sqrt{\frac{(p+q)!}{(p-q)!}} \sqrt{\frac{\theta}{\sin\theta}} J_q(\sqrt{p(p+1)}\theta),
\end{equation}

\begin{equation}
P_p^1(\cos\theta) = -\sqrt{p(p+1)} \sqrt{\frac{\theta}{\sin\theta}} J_1(\sqrt{p(p+1)}\theta).
\label{eq:SmallAngle}
\end{equation}

This small-angle approximation yields numerically faster 2D plots that are not noticeably different from the plot obtained through the use of the exact expression of the associated Legendre functions.

\subsection{Approximations for fields along the z-axis (1D)}
When plotting only the values of the longitudinal intensity profile of the PNJ, we are interested only in the fields along the z-axis (i.e., $\theta=0$); the angular functions (Eqs. \eqref{eq:N}-\eqref{eq:M}) for $\theta \rightarrow 0$ are

\begin{equation}
\lim_{\theta \to 0} \frac{1}{\sin\theta} \left( \tau_p^1(\theta) \cos\theta - \pi_p^1(\theta) \right) = 0,
\end{equation}

\begin{equation}
\lim_{\theta \to 0} \frac{\partial \tau_p^1(\theta)}{\partial \theta} = 0,
\end{equation}

\begin{equation}
\lim_{\theta \to 0} \tau_p^1(\theta) =-\frac{1}{2}p(p+1),
\end{equation}

\begin{equation}
\lim_{\theta \to 0} \pi^1(\theta) =-\frac{1}{2}p(p+1).
\end{equation}

This greatly simplifies the field, especially for a fast calculation of the longitudinal profile using the Poynting vector of section \ref{sec:PoyntingVector}, resulting in fields with no $r$-component.

\subsection{Time-averaged Poynting vector}
\label{sec:PoyntingVector}
By examining the time-averaged Poynting vector, we aim to elucidate the vectorial characteristics of the PNJ and its dependence on the incident polarization. Previous results \cite{Lecler2005} considered the total field for scattering by a microsphere to represent the PNJ. Here, emphasis is given to the extinction component of the Poynting vector \cite{Berg2014} as the origin of the PNJ, as suggested in \cite{Devilez2009}. We can now determine the Poynting vector for the incident, scattered, and interference fields. The electromagnetic fields outside the spherical scatterer are
\begin{equation}
\vec{E}_{tot} = \vec{E}_{inc}+ \vec{E}_{sca}, \quad \vec{H}_{tot} = \vec{H}_{inc}+ \vec{H}_{sca}.
\end{equation}
The time-averaged Poynting vectors are as follows:
\begin{align}
\big \langle \vec{S}_{inc} \big \rangle &= \frac{1}{2} \Re \left( \vec{E}_{inc} \times \vec{H}_{inc}^* \right), \\
\big \langle \vec{S}_{sca} \big \rangle &= \frac{1}{2} \Re \left( \vec{E}_{sca} \times \vec{H}_{sca}^* \right), \\
\big \langle \vec{S}_{ext} \big \rangle &= \frac{1}{2} \Re \left( \vec{E}_{tot} \times \vec{H}_{tot}^* \right) = \frac{1}{2} \Re \left( \vec{E}_{inc} \times \vec{H}_{sca}^* \right)+\frac{1}{2} \Re \left( \vec{E}_{sca} \times \vec{H}_{inc}^* \right),
\label{eq:Sext}
\end{align}
where $\vec{S}_{inc}$, $\vec{S}_{sca}$, and $\vec{S}_{ext}$ are, respectively, the incident, scattered, and interference (between the scattered and incident beams) Poynting vectors. We now carefully examine each Poynting vector by using $G_{p}=\sqrt{\pi (2p+1)} i^{p}$ for the case of plane waves. Since by definition, $\left| p_x \right|^2 + \left| p_y \right|^2 = 1$, we recover an on-axis intensity profile independent of the chosen incident polarization as follows:

\begin{equation} 
\big \langle \vec{S}_{inc}\cdot\hat{z} \big \rangle =\frac{1}{8} \left[\sum_{p=1}^{\infty} (2p+1) i^{p} \left( \frac{j_p(kz)}{kz} + \frac{\diff j_p(kz)}{\diff (kz)} -i j_p(kz) \right) \right]^2.
\label{eq:Sinc}
\end{equation}

By using the relations and identities of spherical functions, we recover the Poynting vector for the incident field as

\begin{equation} 
\big \langle \vec{S}_{inc}\cdot\hat{z} \big \rangle = \frac{1}{2} \Re  \left|i e^{ikz} \right|^2=\frac{1}{2}.
\end{equation}

This is expected, as the sum term must converge in Eq. \eqref{eq:Sinc} to a particular value because the incident plane wave is properly defined as

\begin{equation}
\big \langle \vec{S}_{inc} \big \rangle = \frac{1}{2} \Re \left(E_0 H_0^* (p_x\hat{x}+p_y\hat{y}) \times (-p_y^*\hat{x}+p_x^*\hat{y}) \right)= \frac{1}{2} \Re \left(E_0 H_0^* \hat{z} \right).
\end{equation}

This plane-wave analysis validates the present equations. For $S_{sca}$, we start with
\begin{align} 
\big \langle \vec{S}_{sca}\cdot\hat{z} \big \rangle &= \frac{1}{8} \Re \biggl(
\sum_{p=1}^{\infty} (2p+1) i^p \bigg[a_p \left((p+1)\frac{h_p^{(1)}(kz)}{kz} - h_{p+1}^{(1)}(kz)\right)+i b_p h_p^{(1)}(kz) \bigg] \nonumber \\
&\sum_{p=1}^{\infty} (2p+1) (-i)^p\bigg[b_p^* \left((p+1)\frac{h_p^{(1)*}(kz)}{kz} - h_{p+1}^{(1)*}(kz)\right)-i a_p^* h_p^{(1)*}(kz) \bigg] \biggl).
\end{align}
It is tempting to approximate the spherical Hankel function in its asymptotic form, since the fields are in the region outside the spherical scatterer, i.e., $kr>ka$. For a low $p$-index, $kr>>p$. However, when $p$ increases, the Mie coefficient decreases; in such a case, we expect the asymptotic form to be a very good approximation. Unfortunately the asymptotic expansion does not represent the Poynting vector up to twice the size factor.

Finally, we are interested in determining the Poynting vector related to the interference between the incident and scattered beams. From Eq. \eqref{eq:Sext}, this would be the real part of a linear combination of the spherical Hankel function with a plane wave, which is described in the form of the equation below:

\begin{equation} 
\big \langle \vec{S}_{ext}\cdot\hat{z} \big \rangle = \frac{1}{4} \Re \biggl[i e^{ikz} \sum_{p=0}^{\infty} (-i)^p h_p^{(1)*}(kz) B_p  - i e^{-ikz} \sum_{p=0}^{\infty} i^p h_p^{(1)}(kz) A_p \biggl].
\label{eq:SextLong}
\end{equation}

Note that $A_p$ and $B_p$ are linear combinations of the complex Mie coefficients $a_p$ and $b_p$, and the position dependence of the Poynting vector originates from the terms containing ($kz$) in Eq. \eqref{eq:SextLong}.

\subsection{Optical forces}
\label{sec:OpticalForces}

Once the BSCs are calculated, the force components on an optically trapped microsphere are determined \textit{via} the Maxwell stress tensor formalism, providing new insights owing to the analytical nature of the solution. Finally, to represent the PNJ due to an optically trapped microsphere, we first determine the location along the propagation axis where stable equilibrium is verified. The position is determined from an optical-force curve with respect to the trap position from the microsphere center. This type of force profile has been applied previously in the investigation of morphology dependent resonances of an optically trapped microsphere \cite{Neves2006oe}, as well as locations along the axis \cite{Neves2007}. At this stable position, an energy-density plot can be drawn to identify the existence/location of PNJs for different optical trapping parameters (wavelength, refractive indexes, polarization, and numerical aperture).

We know that the optical force along the z-axis is
\begin{align}
F_z&=-\frac{\epsilon \left|E_0\right|^2}{2k^2} \Re \sum_{p=1}^{\infty} \sum_{q=-p}^{+p} \frac{i}{p+1}\biggl\{ \sqrt{\frac{p(p+2)(p+q+1)(p-q+1)}{(2p+3)(2p+1)}} \bigg[ \left(a_{p+1} + a_p^*-2a_{p+1}a_p^*\right) G_{p+1,q}^{TM} G_{p,q}^{TM*} \nonumber \\
&+\left(b_{p+1} + b_p^*-2b_{p+1}b_p^*\right) G_{p+1,q}^{TE} G_{p,q}^{TE*}\bigg] -\frac{q}{p} \left[ \left(a_{p} + b_p^*-2a_p b_p^* \right) G_{p,q}^{TM} G_{p,q}^{TE*} \right] \biggl\}.
\end{align}

For the on-axis case, only $p=\pm1$ remains:
\begin{align}
F_z&=-\frac{\epsilon \left|E_0\right|^2}{2k^2} \Re \sum_{p=1}^{\infty} \frac{i}{p+1}\biggl\{ 
\frac{p(p+2)}{\sqrt{(2p+3)(2p+1)}} \bigg[ \left(a_{p+1} + a_p^*-2a_{p+1}a_p^*\right) \left(G_{p+1,1}^{TM} G_{p,1}^{TM*} + G_{p+1,-1}^{TM} G_{p,-1}^{TM*} \right) \nonumber \\
&+\left(b_{p+1} + b_p^*-2b_{p+1}b_p^*\right) \left( G_{p+1,1}^{TE} G_{p,1}^{TE*} + G_{p+1,-1}^{TE} G_{p,-1}^{TE*} \right) \bigg] -\frac{1}{p} \left[ \left(a_{p} + b_p^*-2a_p b_p^* \right) \left(G_{p,1}^{TM} G_{p,1}^{TE*} -G_{p,-1}^{TM} G_{p,-1}^{TE*}\right)\right] \biggl\}.
\end{align}

The cross products of BSCs for the focused beam can be simplified to

\begin{equation}
\left(G_{p+1,1}^{TM} G_{p,1}^{TM*} + G_{p+1,-1}^{TM} G_{p,-1}^{TM*} \right)=2G_{p+1} G_p^*,
\end{equation}

\begin{equation}
\left(G_{p+1,1}^{TE} G_{p,1}^{TE*} + G_{p+1,-1}^{TE} G_{p,-1}^{TE*} \right)=2G_{p+1} G_p^*,
\end{equation}

\begin{equation}
\left(G_{p,1}^{TM} G_{p,1}^{TE*} - G_{p,-1}^{TM} G_{p,-1}^{TE*} \right)=-2iG_p G_p^*.
\end{equation}

Therefore, the axial force becomes
\begin{align}
F_z&=-\frac{\epsilon \left|E_0\right|^2}{k^2} \Re \sum_{p=1}^{\infty} \frac{i}{p+1}\biggl\{ \frac{p(p+2)G_{p+1} G_p^*}{\sqrt{(2p+3)(2p+1)}}\left( a_{p+1} + a_p^*-2a_{p+1}a_p^*+b_{p+1} + b_p^*-2b_{p+1}b_p^*\right) \nonumber \\
&-\frac{iG_p G_p^*}{p} \left(a_{p} + b_p^*-2a_p b_p^* \right) \biggl\}.
\end{align}

This expression for the axial component of the force must be determined for each position of the beam focus with respect to the sphere center, which involves determining the BSCs for each of these positions. Owing to the linearity of the electromagnetic fields, we can add another focused field simply by adding BSCs. Let us denote, $z_1$ and $z_2$ as the focal positions of the two independent focused beams, instead of the previous notation of $z_0$ for the single beam in Eq. \eqref{eq:BSCfb}. For the double beam the corresponding BSC is

\begin{equation}
G_p^{fb}=i kf e^{ikf} \frac{G_p^{pw}}{p(p+1)} \int_{0}^{\alpha_{max}} \diff\alpha \sin\alpha \sqrt{\cos\alpha} \left( e^{ikz_1\cos\alpha}+e^{ikz_2\cos\alpha}\right) e^{-(f \sin\alpha/\omega)^2}\left[\pi_p^1(\alpha)+ \tau_p^1(\alpha) \right],
\end{equation}

which can be simplified if the beam positions are symmetric (i.e., when $z_1=z_0+d/2$ and $z_2=z_0-d/2$, where $d$ represents the spaceing between both focus) as follows:

\begin{equation}
G_p^{fb} = 2i kf e^{ikf} \frac{G_p^{pw}}{p(p+1)} \int_{0}^{\alpha_{max}} \diff\alpha \sin\alpha \sqrt{\cos\alpha} e^{ikz_0\cos\alpha} \cos\left(kd/2\,\cos\alpha\right) e^{-(f \sin\alpha/\omega)^2}\left[\pi_p^1(\alpha)+ \tau_p^1(\alpha) \right].
\end{equation}

\section{Results and discussion}
\label{sec:ResultsAndDiscussion}

Mathematica (version 10, Wolfram Inc.) was chosen as the computational software for performing the numerical simulation presented in this section because of its arbitrary-precision computation, error tracking, and numerical libraries. A suitable truncation number is chosen to terminate the infinite sum, according to \cite{Neves2012}, and the floating-point machine error is chosen. Note that this is correct only for the total cross section, as has been emphasized recently \cite{Allardice2014}. However, for specific fields instead of cross sections, a similar rule is applicable in which the maximum number of terms greater than the largest $kr$ in the plot region should be chosen. 

\subsection{PNJ as a Bessoidal-type surface}

\begin{figure}[h!]
\centering
\includegraphics[width=10cm]{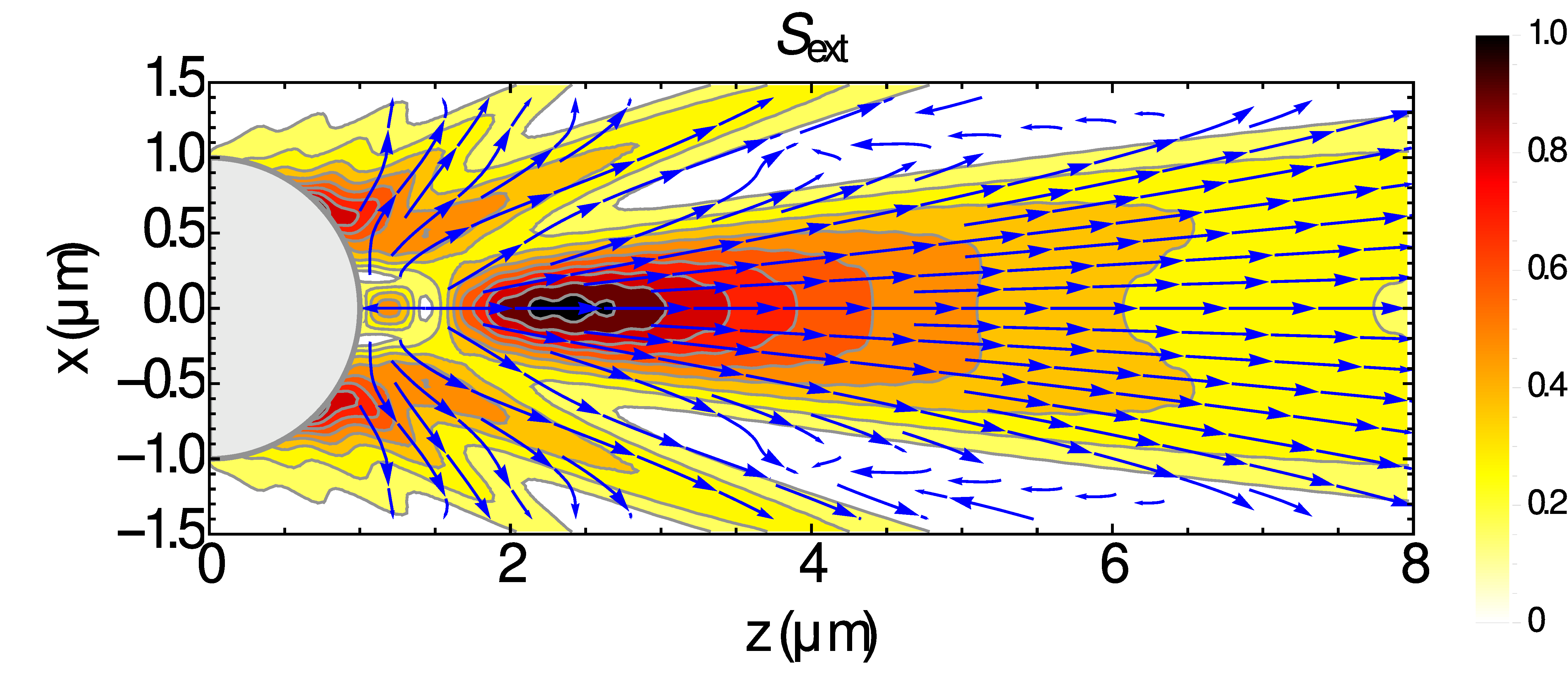}
\caption{Contour and vector plot of the intensity and time-averaged Poynting vector, respectively for the interference between the incident and scattered field. The result is for a 2-$\mu$m polystyrene microsphere ($n$=1.59) in water ($n$=1.33) under plane-wave incidence at $\lambda$=633 nm.}
\label{fig:PW_D2}
\end{figure}

\begin{figure}[h!]
\centering
\includegraphics[width=10cm]{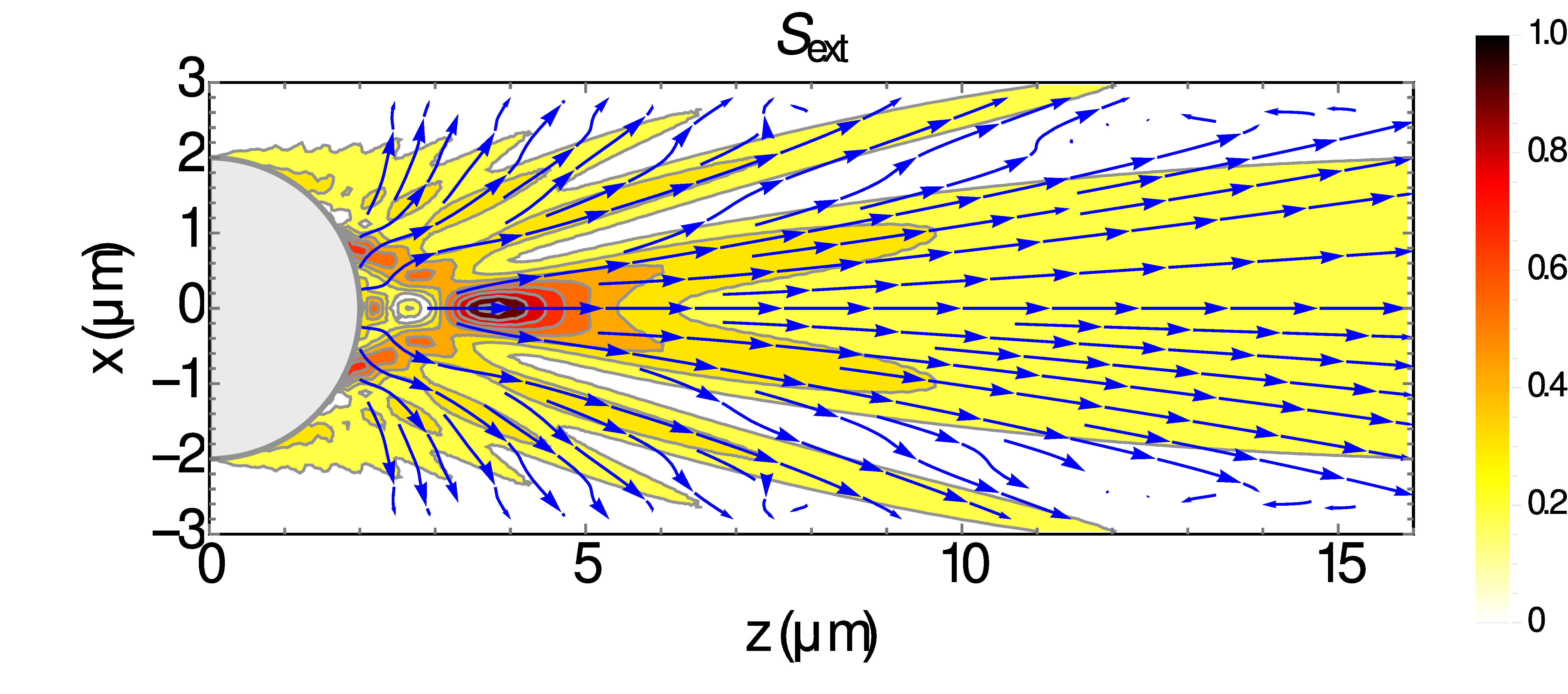}
\caption{Contour and vector plot of the intensity and time-averaged Poynting vector, respectively for the interference between the incident and scattered fields. The result is for a 4-$\mu$m polystyrene microsphere ($n$=1.59) in water ($n$=1.33) under plane-wave incidence at $\lambda$=633 nm.}
\label{fig:PW_D4}
\end{figure}

Starting with conventional plane-wave incidence, we determine the Poynting vector of the interference between the incident and scattered fields in a region immediately after the scatterer for a 2-$\mu$m (Figure \ref{fig:PW_D2}) and 4-$\mu$m microsphere (Figure \ref{fig:PW_D4}).

\begin{figure}[h!]
\centering
\begin{subfigure}{5cm}
  \centering
  \includegraphics[width=.9\linewidth]{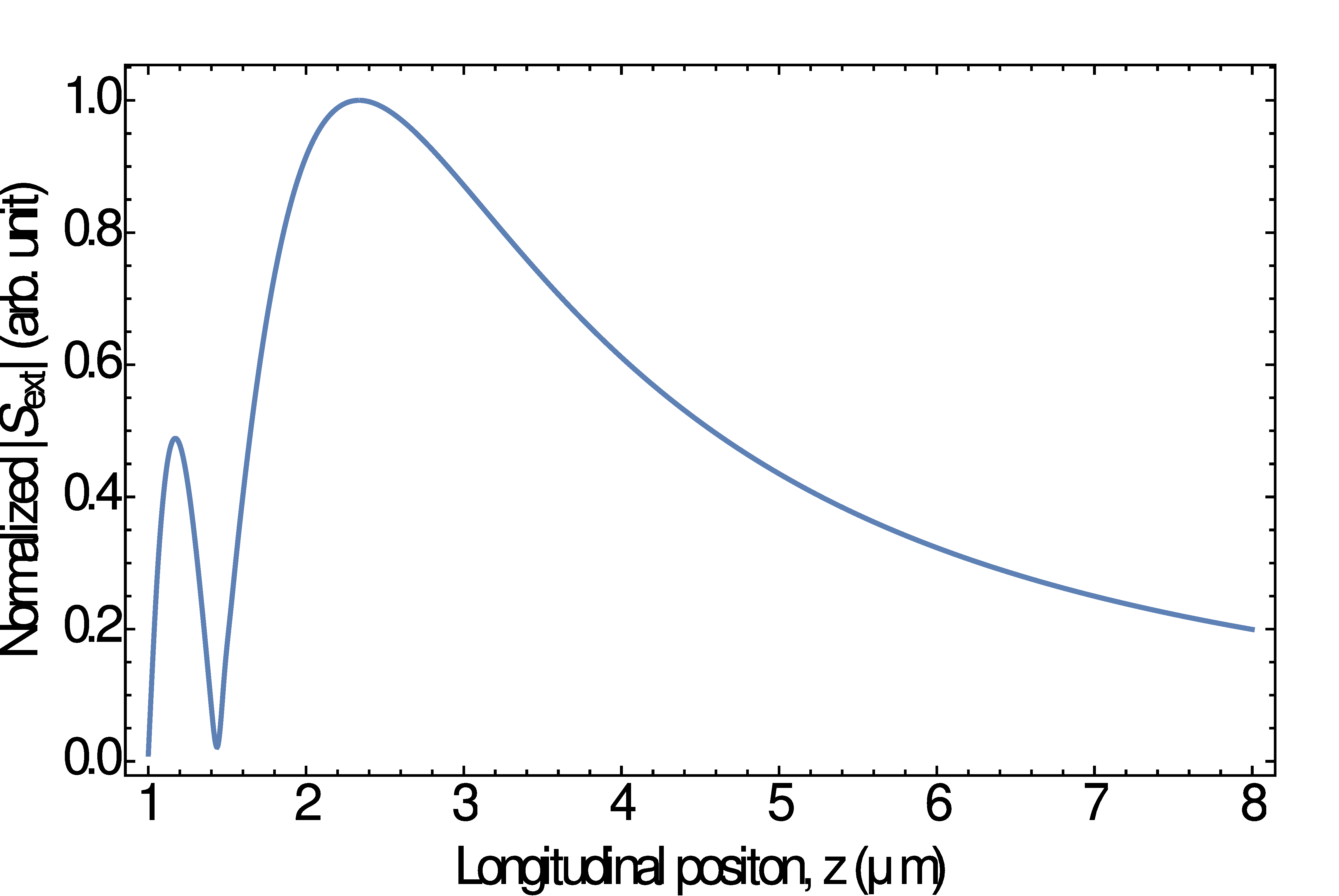}
  \caption{Longitudinal intensity profile}
  \label{fig:sub1}
\end{subfigure}%
\begin{subfigure}{5cm}
  \centering
  \includegraphics[width=.9\linewidth]{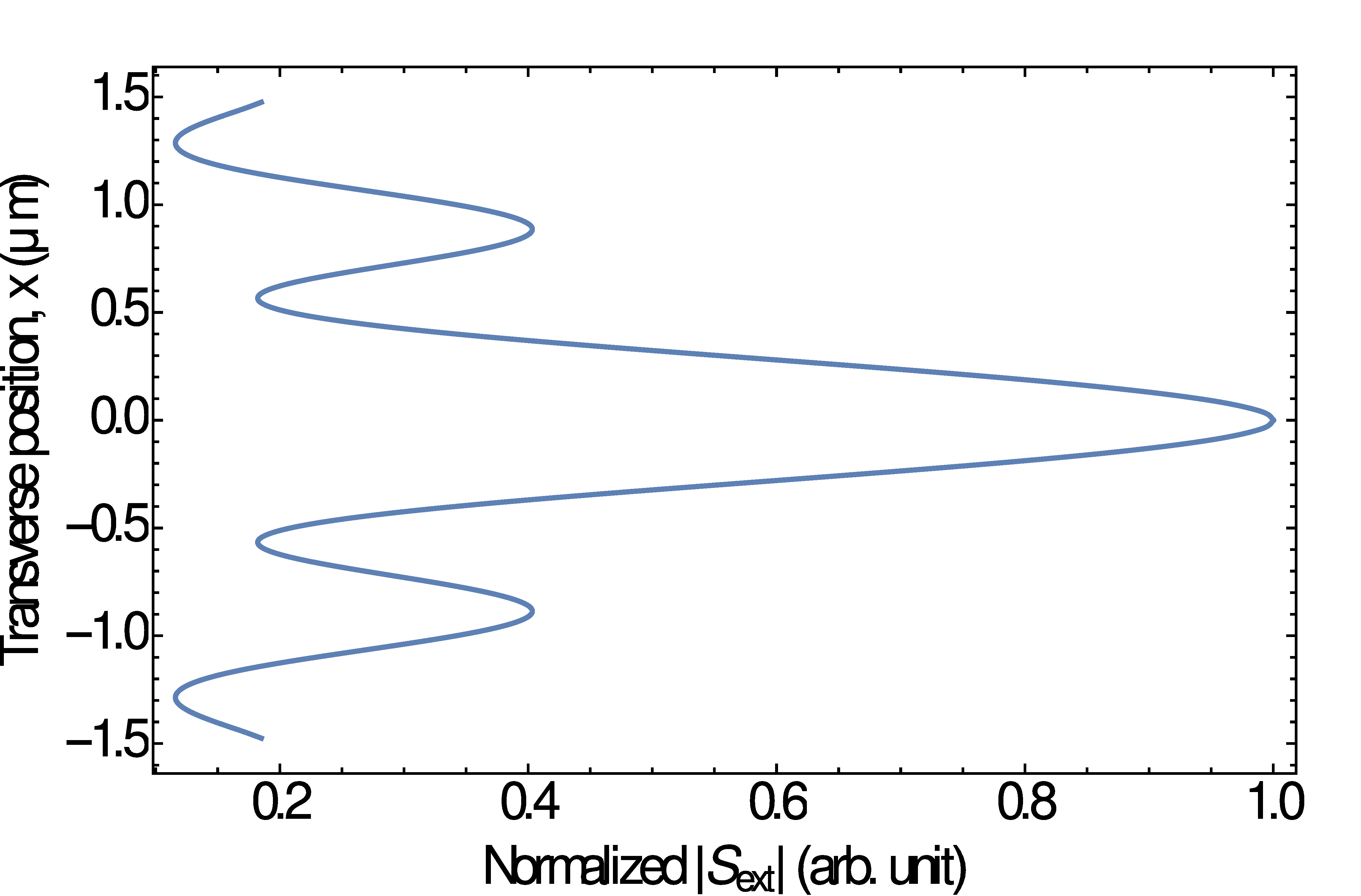}
  \caption{Transverse intensity profile}
  \label{fig:sub2}
\end{subfigure}
\caption{PNJ intensity profile for plane-wave incidence. The result is for the 2-$\mu$m sphere of Figure \ref{fig:PW_D2} at the maximum intensity.}
\label{fig:PW_Profile}
\end{figure}

\begin{figure}[h!]
\centering
\includegraphics[width=8cm]{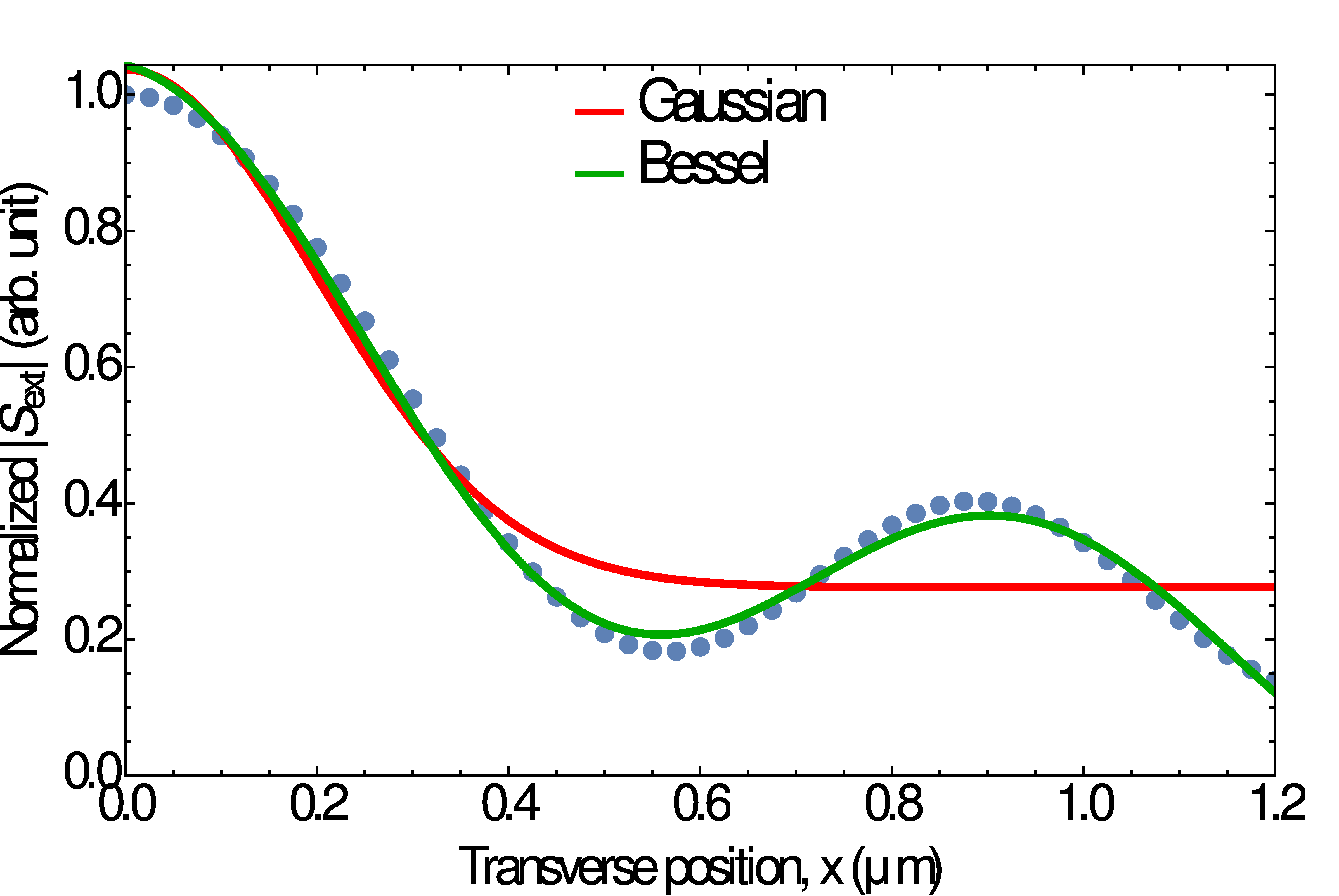}
\caption{Fitting of the PNJ transverse profile using a cylindrical Bessel type versus that using a Gaussian function. The result is for the 2-$\mu$m sphere of Figure \ref{fig:sub2}, clearly showing a better fitting with the Bessel-type function.}
\label{fig:PW_TransverseFitting}
\end{figure}

The fitting function for the transverse and longitudinal intensity profiles of a PNJ has been described as a Gaussian and a Lorentzian, respectively \cite{Devilez2009spie}, noting that the intensity distribution of photonic jets along the z-axis (longitudinal direction) is not symmetric with respect to the maximum intensity. From the results presented here (Figure \ref{fig:PW_Profile}), we can obtain a dominant function that adequately represents the profiles of the PNJ. The transverse profile is better described by spherical (from Eq. \eqref{eq:SextLong}) and cylindrical Bessel functions (from the small-angle approximation, Eq.\eqref{eq:SmallAngle}), as can be observed by the fitting (Figure \ref{fig:PW_TransverseFitting}).

\subsection{PNJ from a focused beam}

\begin{figure}[h!]
\centering
\includegraphics[width=10cm]{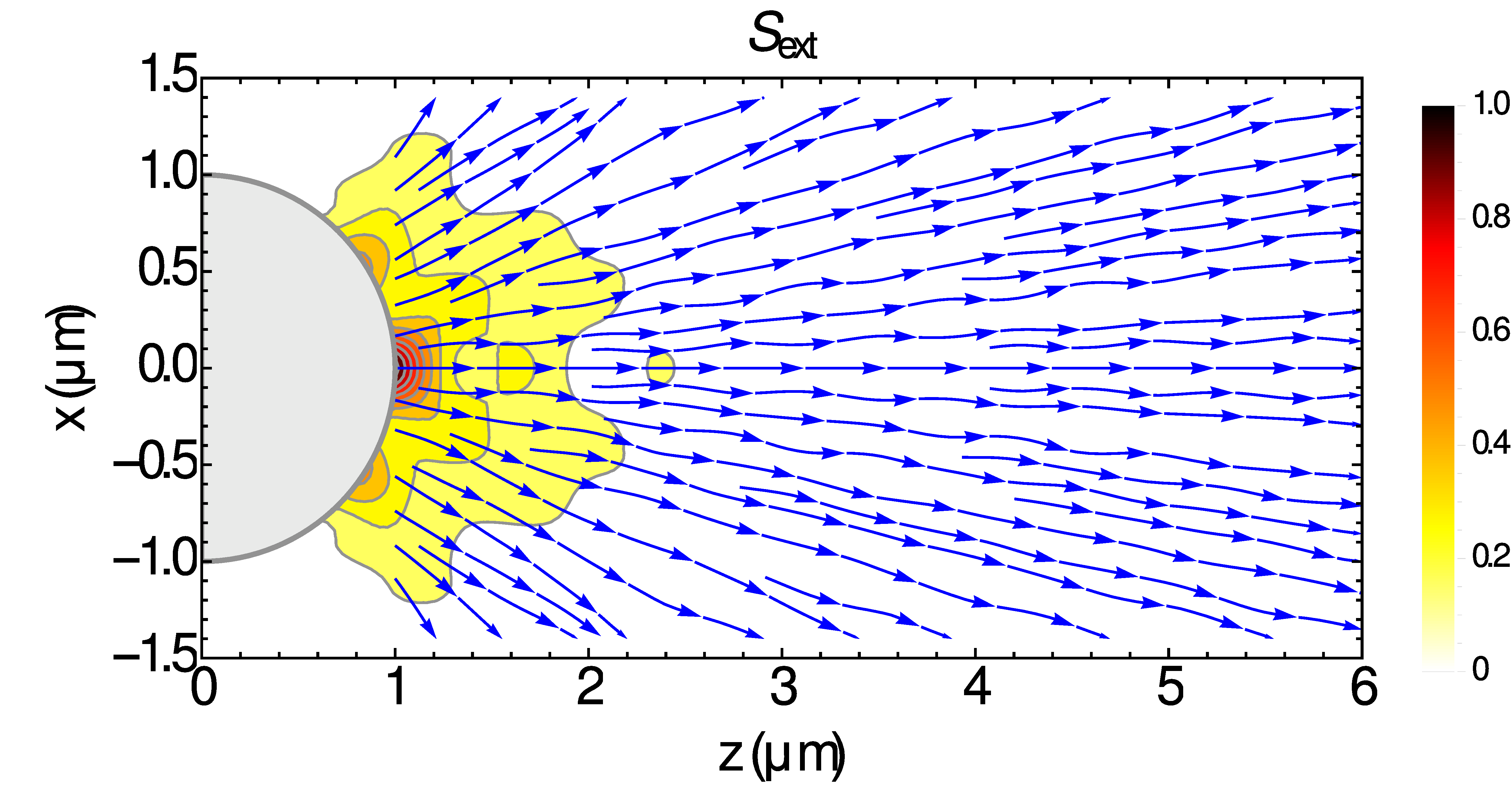}
\caption{Contour and vector plot of the intensity and time-averaged Poynting vector, respectively for the interference between the incident and scattered fields. The result is for the 2-$\mu$m sphere of Figure \ref{fig:sub2}, but with a highly focused ($NA$=1.25) Gaussian beam (waist=2.5mm before at the objective back aperture) placed at the origin ($z_0$ = 0$\mu$m).}
\label{fig:FW_D2Z0}
\end{figure}

A PNJ is only observed when the focus is close to the sphere, as has been pointed out by Lecler \textit{et al.} \cite{Lecler2005}, and the behaviour of the PNJ with respect to the focal position of the beam was investigated by Devilez \textit{et al.} \cite{Devilez2009}. Therefore, for a highly focused beam at the sphere center, which is close to the stable trapping configuration of an optical tweezer, we obtain no PNJ as illustrated in Figure \ref{fig:FW_D2Z0}.

\begin{figure}[h!]
\centering
\includegraphics[width=10cm]{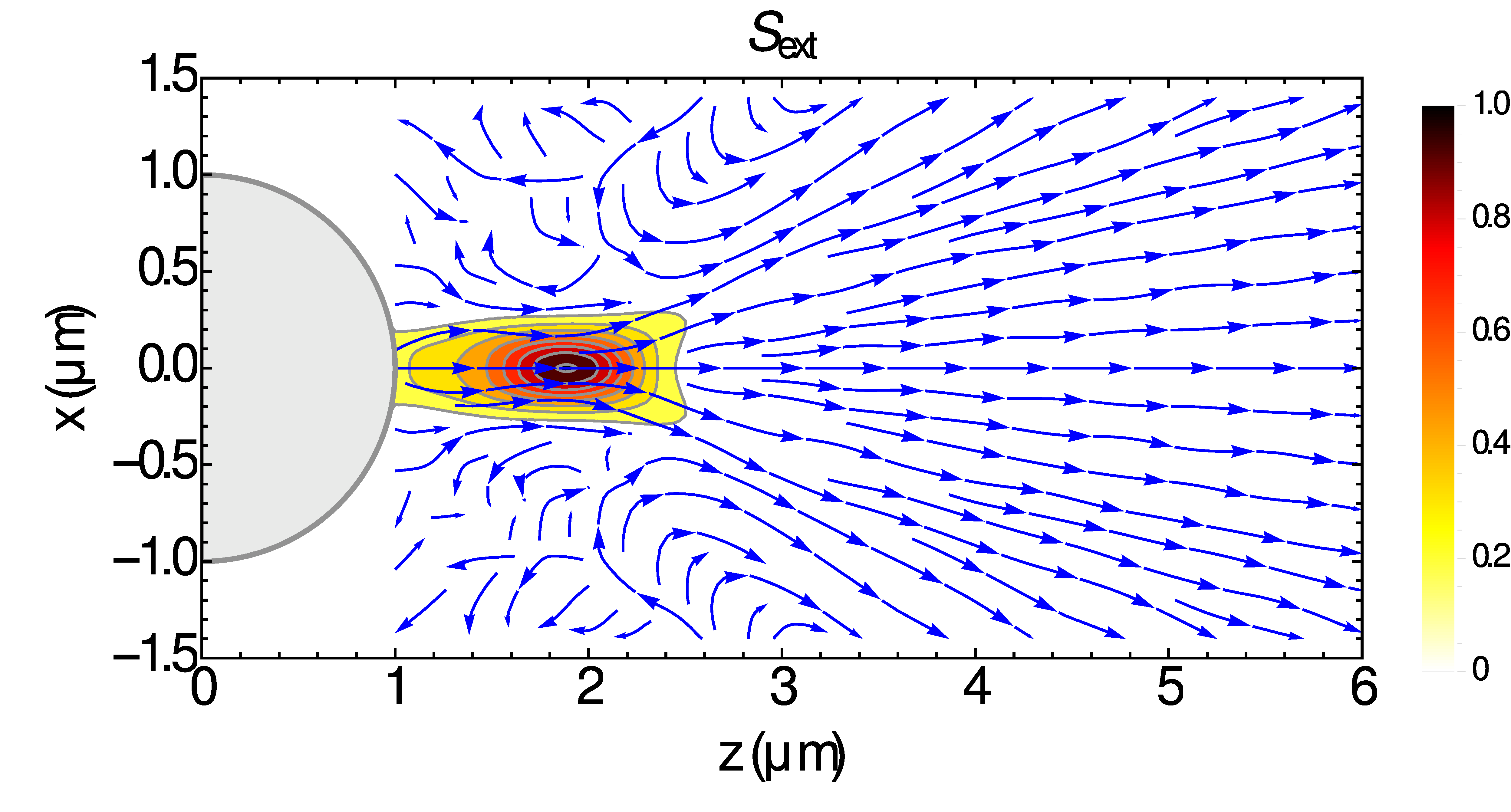}
\caption{Contour and vector plot of the intensity and time-averaged Poynting vector for the interference between the incident and scattered fields. The result is for the 2-$\mu$m sphere of figure \ref{fig:sub2}, but with a highly focused ($NA$=1.25) Gaussian beam (waist=2.5mm before at the objective back aperture) located outside the sphere ($z_0$ = 2$\mu$m).}
\label{fig:FW_D2Z2}
\end{figure}

\begin{figure}[h!]
\centering
\includegraphics[width=8cm]{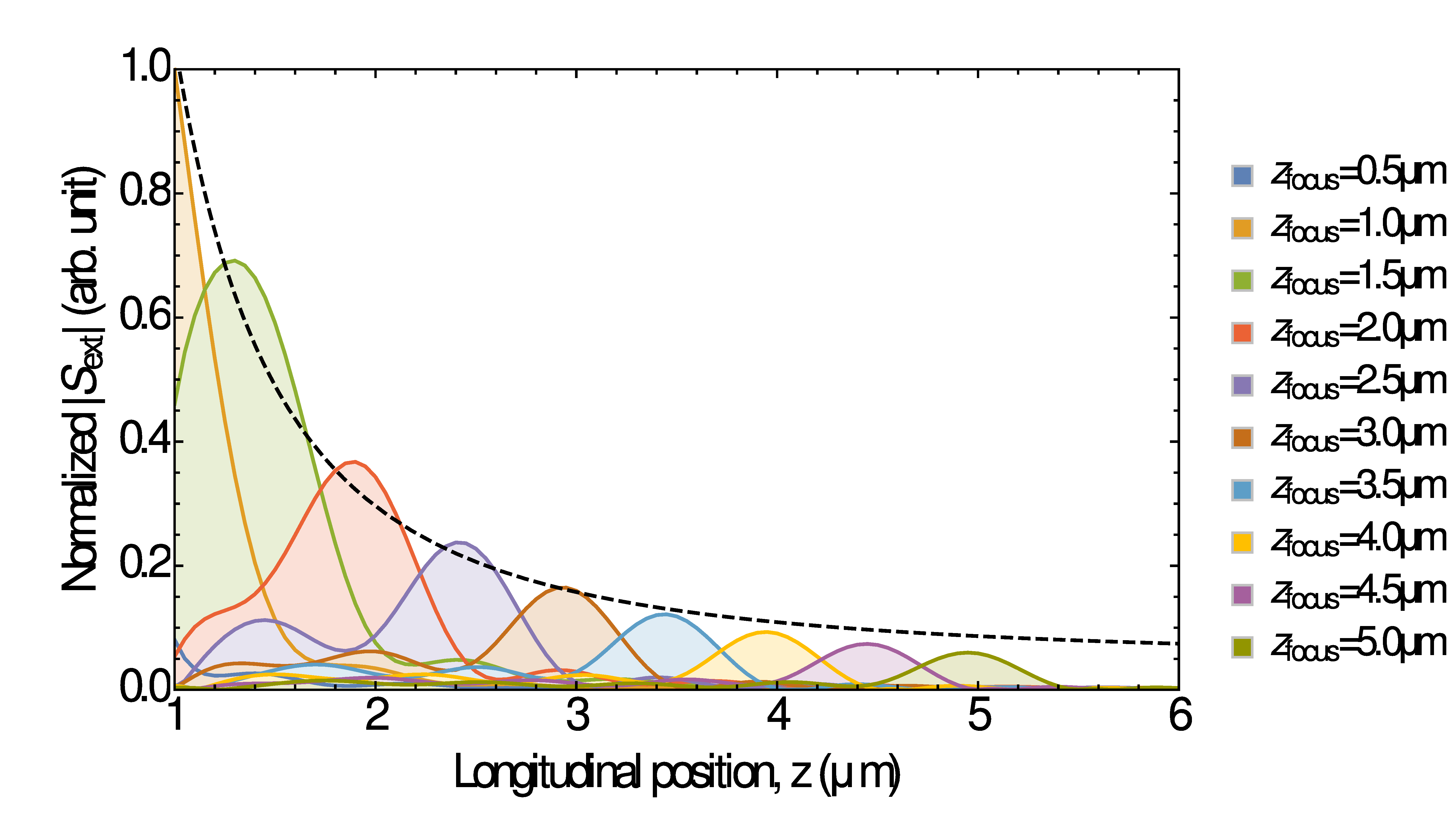}
\caption{Overlapping plot of the on-axis intensity of the time-averaged Poynting vector for the interference between the incident and scattered fields. The result is for the 2-$\mu$m sphere of figure \ref{fig:sub2}, but with a highly focused ($NA$=1.25) Gaussian beam (waist=2.5mm before at the objective back aperture) located at specific positions ($z_0$ = $z_{focus}$), i.e., different radial shifts. The dashed line is an $r^{-2}$ fitting though the maximum profile points.}
\label{fig:FW_D2FocalPositions}
\end{figure}

To observe any appreciable PNJ, we have to shift the incident-beam focus close to the surface. By placing it 2 $\mu$m from the sphere center, we observed the PNJ of Figure \ref{fig:FW_D2Z2}. Moreover, we examined the PNJ transverse profile as a function of beam position, as shown in Figure \ref{fig:FW_D2FocalPositions}. It can be observed that the PNJ due to the interference of the incident and scattered fields, inherits the $r^{-2}$ intensity-decay characteristic of the scattered field and localization from the highest-intensity position of the incident field.

The challenge here is to obtain a PNJ from an optically trapped sphere. In an optical tweezer, we should not have a PNJ, since the microsphere center is brought near the focus of the beam, in a configuration similar to that of Figure \ref{fig:FW_D2Z0}.

\subsection{Optical forces in a microsphere}
\begin{figure}[h!]
\centering
\includegraphics[width=10cm]{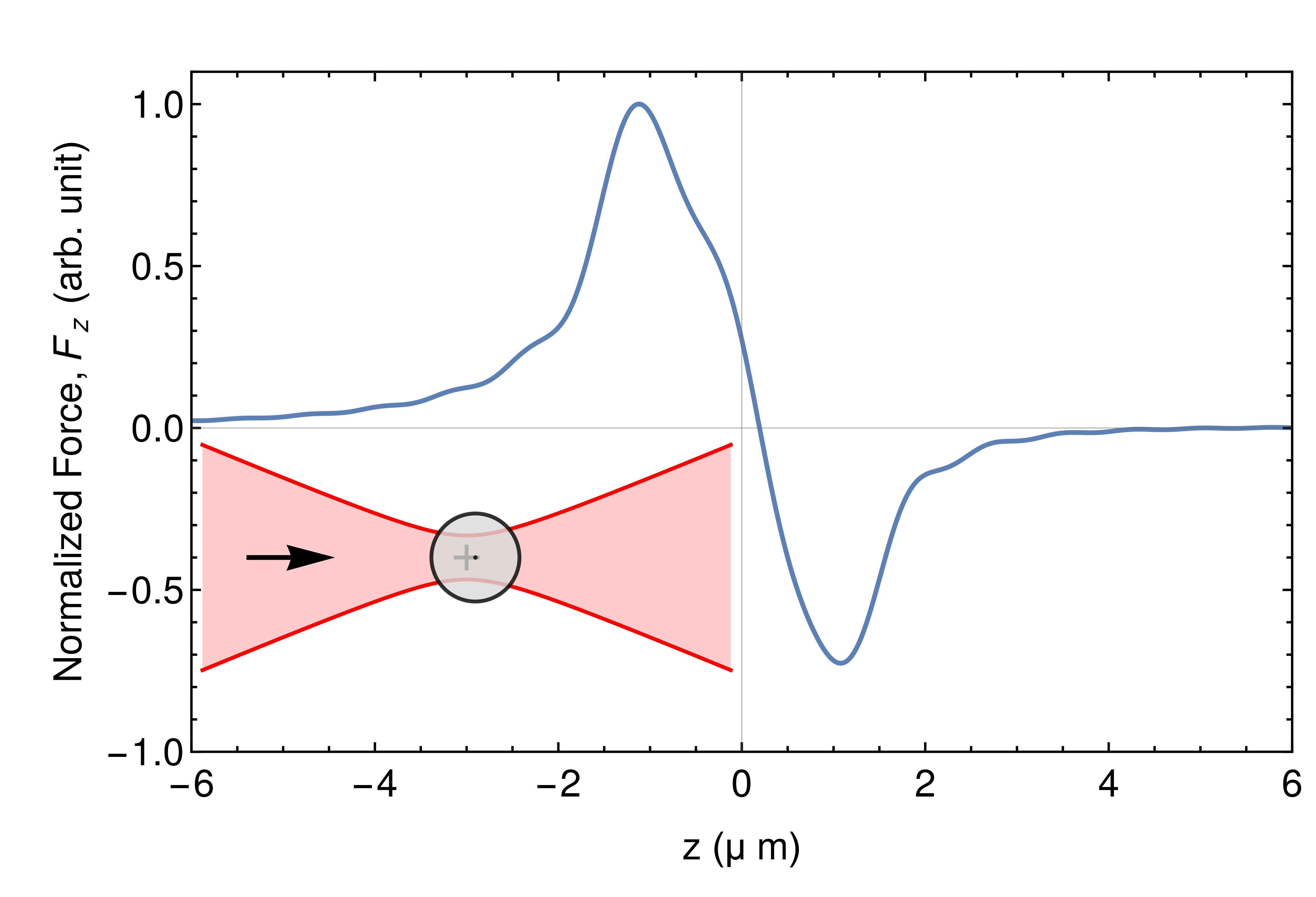}
\caption{Axial optical force as a function of microsphere displacement $z (\mu m)$ from the focus. The inset illustrates equilibrium configuration, the beam propagation direction (arrow), the focus (plus) and microsphere (dot).}
\label{fig:OT_ForceZ}
\end{figure}

The optical force profile of conventional optical tweezers is presented in Figure \ref{fig:OT_ForceZ}. Note that the equilibrium position for the microsphere is located immediately after ($\approx$ 0.2 $\mu$m) the highly focused beam position, which is due to the scattering forces as the focused beam close to the sphere center yields no PNJ.

To circumvent the absence of a PNJ in conventional optical tweezers, we adopt double optical tweezers: not the configuration commonly employed in which each beam axis is parallel to each other \cite{Fontes2005}, but in a configuration in which both beams are on the same axis. This leads to the assumption of the original counter-propagating traps of 1970 \cite{Ashkin1970}, but here, the two traps are co-propagating. In this double co-propagating optical-tweezers configuration, an equilibrium position will exist if the gradient force from one tweezer exceeds the scattering force from the other.

\begin{figure}[h!]
\centering
\begin{subfigure}{6cm}
  \centering
  \includegraphics[width=.9\linewidth]{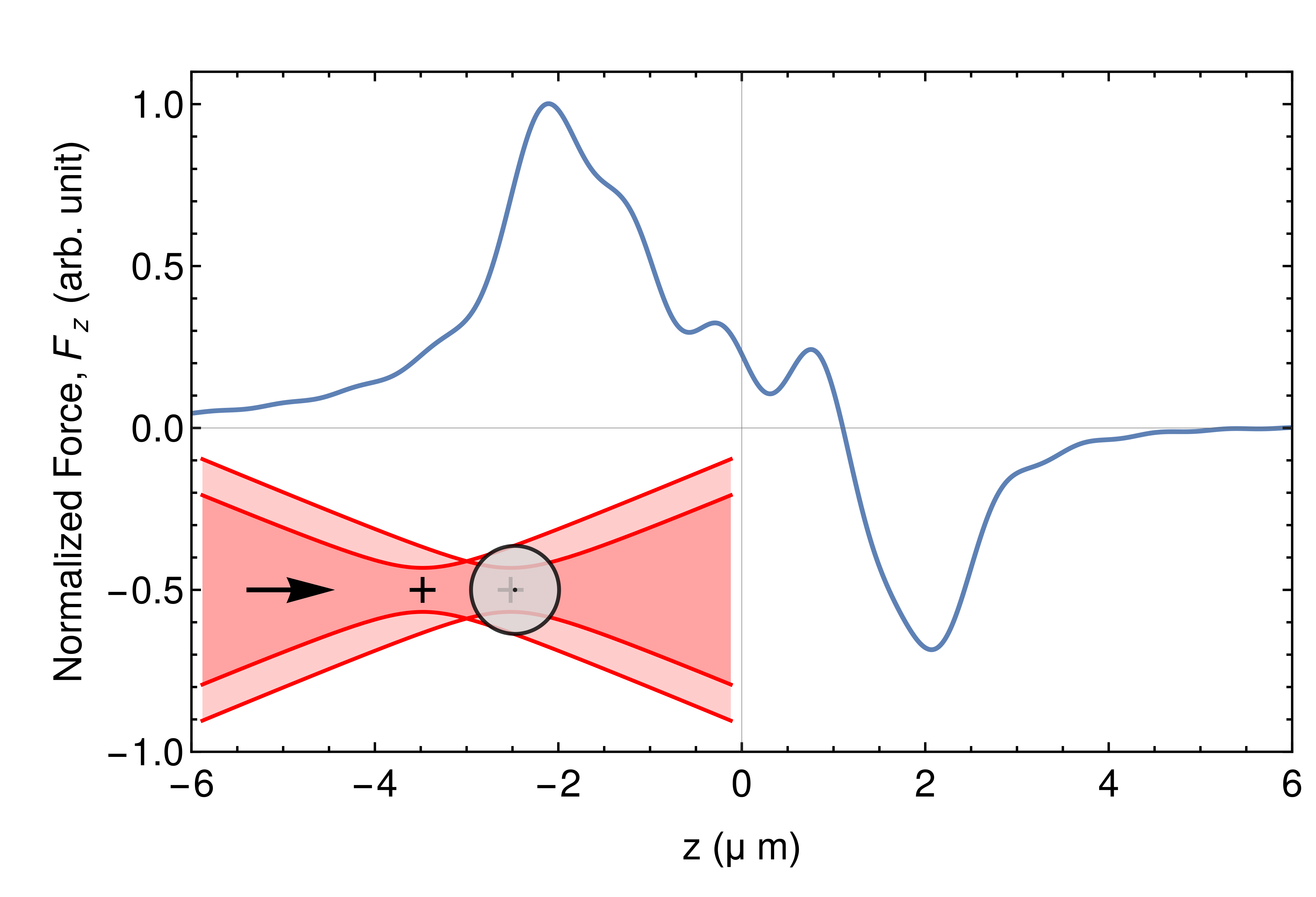}
  \caption{Axial optical-force profile}
  \label{fig:OTsub1}
\end{subfigure}%
\begin{subfigure}{6cm}
  \centering
  \includegraphics[width=.9\linewidth]{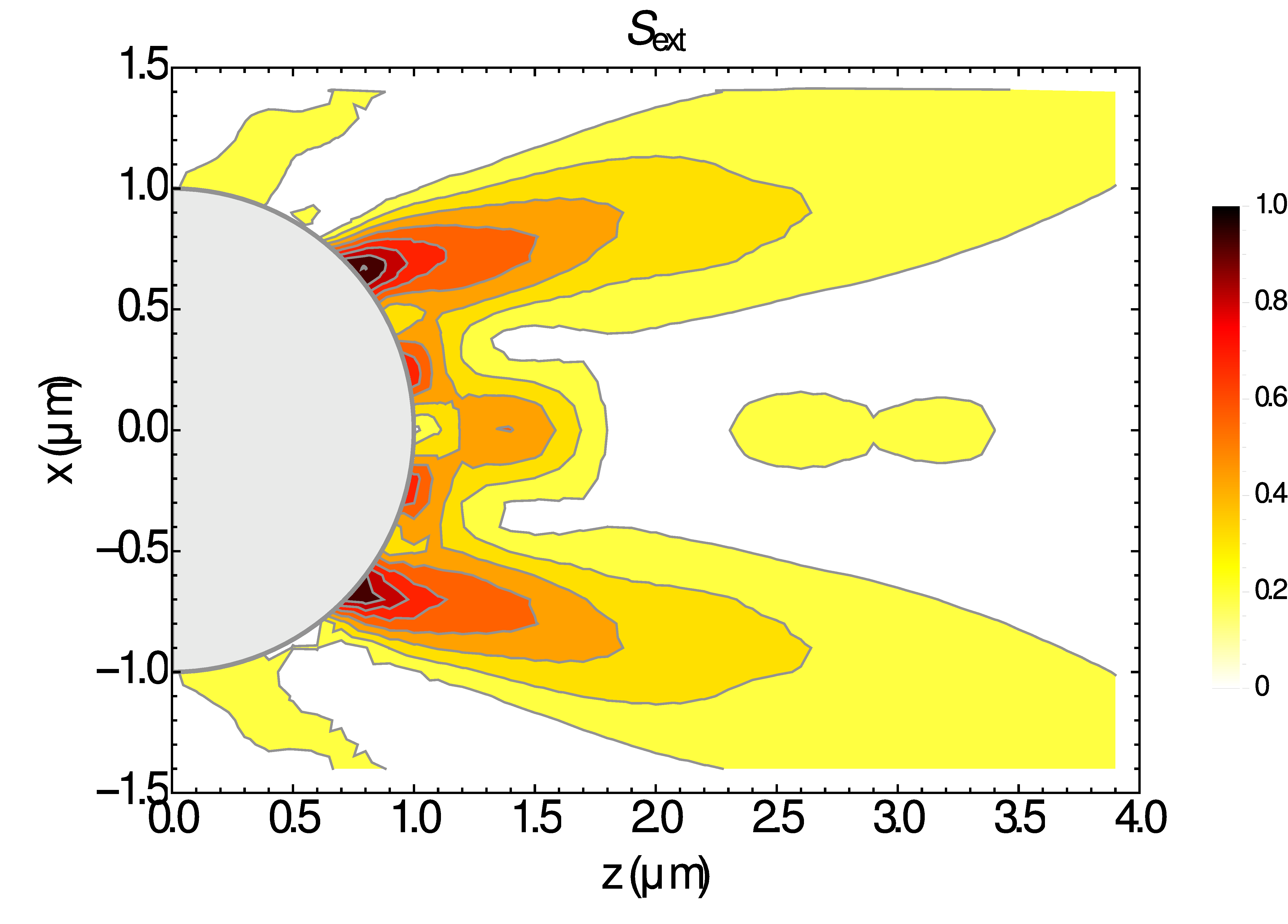}
  \caption{Contour intensity plot}
  \label{fig:Profilesub1}
\end{subfigure}
\begin{subfigure}{6cm}
  \centering
  \includegraphics[width=.9\linewidth]{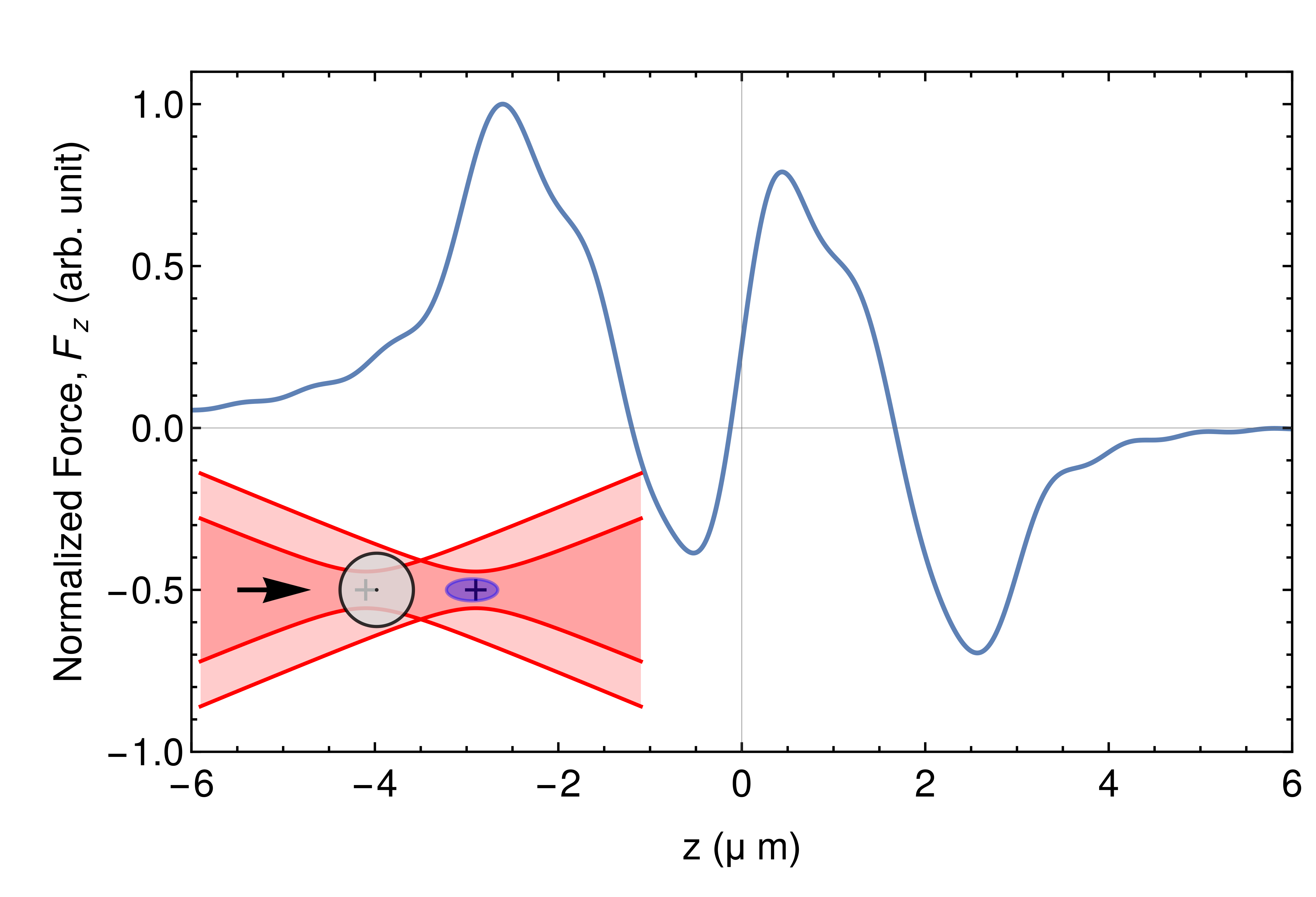}
  \caption{Axial optical-force profile}
  \label{fig:OTsub2}
\end{subfigure}%
\begin{subfigure}{6cm}
  \centering
  \includegraphics[width=.9\linewidth]{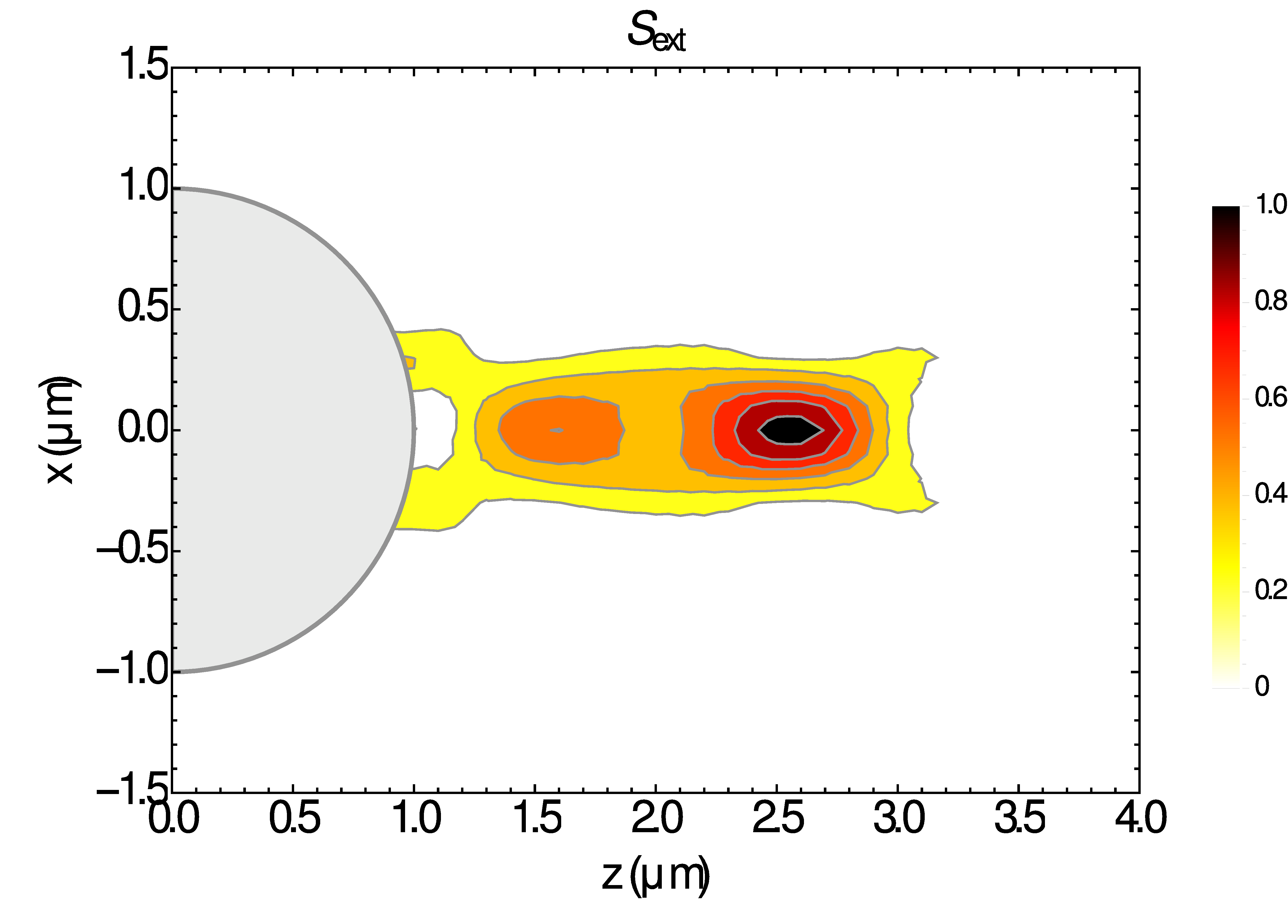}
  \caption{Contour intensity plot}
  \label{fig:Profilesub2}
\end{subfigure}
\caption{Double co-propagating optical traps: one located at z=-1.0 $\mu$m and z=1.0 $\mu$m, corresponding to (a)-(b), and the other located at z=-1.5 $\mu$m and z=1.5 $\mu$m, shown by (c)-(d). The inset illustrates equilibrium configuration, the beam propagation direction (arrow), the focus (plus) and microsphere (dot).}
\label{fig:DOT_Profile_Instensity}
\end{figure}

For the 2-$\mu$m sphere and two equal-intensity traps located 2 $\mu$m apart (Figure \ref{fig:OTsub1}), because of the scattering force from the first trap, the equilibrium position is displaced to about 1.1 $\mu$m in the beam-propagation direction. There is a cost associated with using these co-propagating traps: the reduction of the trap stiffness. The loss of stiffness for the present case is approximately 40$\%$ with respect to a single optical tweezer. Consequently, the second focused beam is positioned within the trapped sphere, and no PNJ is generated (Figure \ref{fig:Profilesub1}). On increasing the displacement between the two traps to 3 $\mu$m, we obtain two equilibrium positions, as illustrated in Figure \ref{fig:OTsub2}. For the microsphere located in the first equilibrium position of z=-1.2 $\mu$m, a focused beam is present immediately outside the microsphere, resulting in a PNJ (Figure \ref{fig:Profilesub2}). For the second equilibrium position, there is no PNJ, as expected. Therefore depending on where the microsphere is located within the two possible trap locations, the PNJ can easily be switched on or off by modulating the intensity of both traps.

\section{Conclusion}
\label{sec:Conclusion}
In summary, we presented a detailed theoretical investigation of the formal relationship between the optical force on a microsphere and its photonic nanojet, which results from the interference between the scattered and incident fields. This approach is general, and a complex refractive index can be implemented to take into account light absorption by the scatterer. Even though double optical tweezers were employed, a similar principle holds as for other beam configurations, especially those generated by holographic traps.

Consequently, an understanding between the interaction of an optically trapped microsphere and its PNJ field structure is of fundamental importance in optical physics and has practical significance in applications such as imaging, nano-lithography, detection, metrology, biophotonics, and spectroscopy. Such an understanding also provides an opportunity to manipulate the intensities of two traps differently in order to achieve a controlled PNJ near an optically trapped microsphere. Moreover, the loss in trap stiffness can be easily overcome using anti-reflection coatings to enhance the trapping force \cite{Hu2008}. In the case of trapping soft spheres, such as water drops, as has been recently reported \cite{Wang2014}, laser-induced surface stress would break the spherical symmetry of the scatterer leading to spheroidal particles. The PNJs generated by such scatterers have recently been reported in \cite{Han2014}. Finally when whispering-gallery modes or pulsed beams are of interest, they can also be detected in an optical trap \cite{Fontes2005}, thereby validating recent simulations on photonic-jet shaping \cite{Geints2013a} and  temporal dynamics of the jet \cite{Geints2013b}.

\section*{Acknowledgments}
\label{sec:Acknowledgments}
The present work received support from CNPq (308627/2012-1), Conselho Nacional de Desenvolvimento Cient\'ifico e Tecnol\'ogico and FAPESP (2014/07191-9), Funda\c{c}\~ao de Amparo \`a Pesquisa do Estado de S\~ao Paulo, Brazil.

\appendix
\section{Shape-coefficient integrals of the beam}
\label{Sec:AppIntegrals}

In this appendix, the analytical integrals for the integration of the BSCs are presented. The first integral, in terms of the azimuthal angle $\phi$ is of the type

\begin{equation}
\int_{0}^{2\pi} d\phi e^{-i m \phi} \left[ \begin{array}{c} \cos\phi \\ \sin\phi \end{array} \right]=\pi \left[ \begin{array}{c} \delta_{q,1} + \delta_{q,-1} \\ -i(\delta_{q,1} - \delta_{q,-1}) \end{array} \right],
\end{equation}

which does not depend on $\theta$. Therefore the remaining integral, in terms of the polar angle $\theta$ for the plane-wave case, is of the type

\begin{equation}
\label{integralnova}
\int_{0}^{\pi} d\theta \sin^2\theta \, P_p^q(\cos\theta) e^{i k r \cos\theta}.
\end{equation}

This integral resembles a formerly reported one \cite{Neves2006jpa}, the solution of which is

\begin{equation}
\int_{0}^{\pi} d\theta \sin\theta P_p^q(\cos\theta) e^{i kr \cos\alpha \cos\theta} J_q(kr \sin\alpha \sin\theta) =2 i^{p-q} P_p^q(\cos\alpha)j_p(kr).
\end{equation}

Now, we are interested in taking the limit $\alpha \rightarrow 0$ of the above integral. Because of the properties of the Bessel function and associated Legendre functions, we obtain, for $q=+1$ and  $q=-1$,

\begin{equation}
\int_{0}^{\pi} d\theta \sin^2\theta P_p^{1}(\cos\theta) e^{i kr \cos\theta} = 2 p(p+1) i^{p+1} \frac{j_p(kr)}{kr},
\end{equation}

\begin{equation}
\int_{0}^{\pi} d\theta \sin^2\theta P_p^{-1}(\cos\theta) e^{i kr \cos\theta} = -2 i^{p+1} \frac{j_p(kr)}{kr},
\end{equation}

thereby eliminating the radial-dependency term, $j_p(kr)/kr$, from the BSCs.

\bibliographystyle{elsarticle-num}

\bibliography{NevesArxiv}

\end{document}